\documentclass[twocolumn]{IEEEtran}

\usepackage[pdftex]{graphicx} % 插图命令 \includegraphics
\usepackage{amsmath} % Math Packages
\usepackage{algorithmic}
\usepackage{array} % Alignment Packages
\usepackage[caption=false,font=footnotesize]{subfig} % Subfigure Packages
\usepackage{dblfloatfix}
\usepackage{url} % simple URL typesetting

\usepackage{microtype}
\usepackage[utf8]{inputenc} % allow utf-8 input
\usepackage[T1]{fontenc}    % use 8-bit T1 fonts
\usepackage{paralist} % Compacted versions of enumerate and itemize.
\usepackage{marvosym} % 常用符号, 特别是打勾、打叉
\usepackage{bm} % 实现字体的斜体加粗
\usepackage{bbm} % fix \mathbb doesn't support digits
\usepackage{amssymb}   % 黑板粗体 \mathbb, 德文尖角体 \mathfrak, \backsim
\usepackage{mathrsfs} % 花体 \mathscr
\usepackage{tabularray} % tblr or talltblr or longtblr environment
\UseTblrLibrary{booktabs}       % professional-quality tables，也就是三线表
\usepackage{tabularx} % 自动计算列宽的表格包
\usepackage[flushleft]{threeparttable} % 给表格项添加注释
\usepackage{colortbl} % 表格行列颜色
\usepackage{makecell} % 表格内容换行
\usepackage{multirow} % 多行表格
\usepackage{nicefrac} % 斜着的分号
\usepackage[dvipsnames]{xcolor} % 字体颜色
\usepackage{soul} % highlight, strikeout, underline
\setul{0.5ex}{0.3ex}

\definecolor{citecolor}{RGB}{34,139,34}
\usepackage[pagebackref=true,breaklinks=true,letterpaper=true,colorlinks,
    citecolor=citecolor,bookmarks=false]{hyperref}
\usepackage{cleveref}  % for \cref
\crefformat{figure}{Fig.~#2#1#3}
\crefformat{equation}{Eq.~(#2#1#3)}
\crefformat{table}{Table~#2#1#3}
\crefformat{section}{Section~#2#1#3}
\crefformat{algorithm}{Algorithm~#2#1#3}
\crefformat{appendix}{Appendix #2#1#3}
\crefformat{subsection}{subsection~#2#1#3}
\usepackage{cellspace}
\usepackage[numbers,sort&compress]{natbib} % 代替 cite.sty

\usepackage{xspace} % 用于在 \newcommand 中添加空格

\usepackage{pifont} % http://ctan.org/pkg/pifont
\definecolor{Gray}{rgb}{0.9,0.9,0.9}
\definecolor{LightCyan}{rgb}{0.88,1,1}
\newcolumntype{a}{>{\columncolor{Gray}}c}
\newcolumntype{b}{>{\columncolor{white}}c}

\begin{document}

\setlength{\abovedisplayskip}{.5\baselineskip} % 调整公式与正文间的段前距离
\setlength{\belowdisplayskip}{.5\baselineskip} % 调整公式与正文间的段后距离

% 模板已使用 subfig 宏包，这里移除 subcaption 宏包以避免冲突
% \usepackage{subcaption}

\title{Selective Variable Convolution Meets Dynamic Content-Guided Attention for Infrared Small Target Detection}

% !TEX root = ../main.tex

\author{
  Yirui~Chen,
  Yiming~Zhu,
  Yuxin~Jing,
  Tianpei~Zhang,
  Jufeng~Zhao
  % \thanks{
  %   This work was supported by
  %     % 国家自然科学基金
  %     the National Natural Science Foundation of China (62301261, % 我
  %       62206134, % 翔哥
  %       U24A20330, 62361166670). % 我% 杨健老师
  %   % 博后面上
  %   % the Fellowship of China Postdoctoral Science Foundation (No. 
  %   % 我
  %   % 2021M701727).
  %   \emph{(Corresponding author:
  %    Yimian Dai).}
  %   }

  % % 南理工
  % % \thanks{
  % %   Yimian Dai and Jian Yang are with PCA Lab, Key Lab of Intelligent Perception and Systems for High-Dimensional Information of Ministry of Education, and Jiangsu Key Lab of Image and Video Understanding for Social Security, School of Computer Science and Engineering, Nanjing University of Science and Technology, Nanjing, China.
  % %   (e-mail:
  % %   \href{mailto:yimian.dai@gmail.com}{yimian.dai@gmail.com};
  % %   \href{mailto:csjyang@mail.njust.edu.cn}{csjyang@mail.njust.edu.cn}).
  % % }

  % % 南开
  % \thanks{
  % Yimian Dai, Xiang Li, and Jian Yang are with VCIP, College of Computer Science, Nankai University. 
  % Xiang Li also holds a position at the NKIARI, Shenzhen Futian. 
  % (e-mail:
  % \href{mailto:yimian.dai@gmail.com}{yimian.dai@gmail.com};
  % \href{mailto:xiang.li.implus@nankai.edu.cn}{xiang.li.implus@nankai.edu.cn};
  % \href{mailto:csjyang@nankai.edu.cn}{csjyang@nankai.edu.cn}).
  % }

}

\maketitle

% !TEX root = ../main.tex

\begin{abstract}
	Infrared Small Target Detection (IRSTD) system aims to identify small targets in  complex backgrounds. Due to the convolution operation in Convolutional Neural Networks (CNNs), applying traditional CNNs to IRSTD presents challenges, since the feature extraction of small targets is often insufficient, resulting in the loss of critical features. To address these issues, we propose a dynamic content-guided attention multiscale feature aggregation network (DCGANet), which adheres to the  attention principle of 'coarse-to-fine' and achieves high detection accuracy. First, we propose a selective variable convolution (SVC) module that integrates the benefits of standard convolution, irregular deformable convolution, and multi-rate dilated convolution. This module is designed to expand the receptive field and enhance non-local features, thereby effectively improving the discrimination between targets and backgrounds. Second, the core component of DCGANet is a two-stage content-guided attention module. This module employs a two-stage attention mechanism to initially direct the network's focus to salient regions within the feature maps and subsequently determine whether these regions correspond to targets or background interference. By retaining the most significant responses, this mechanism effectively suppresses false alarms. Additionally, we propose an Adaptive Dynamic Feature Fusion (ADFF) module to substitute for static feature cascading. This dynamic feature fusion strategy enables DCGANet to adaptively integrate contextual features, thereby enhancing its ability to discriminate true targets from false alarms. DCGANet has achieved new benchmarks across multiple datasets.

\end{abstract}

\vspace{-1\baselineskip}

\section{Introduction} \label{sec:introduction}

Infrared imaging technology, a crucial component of infrared monitoring systems, excels at capturing thermal signatures of targets and demonstrates robust performance in low-light and foggy environments. In military \cite{liu2023infrared}, ocean rescue \cite{hu2024smpisd}, and environmental monitoring \cite{kou2023infrared, li2024visible} applications, targets of interest at long imaging distances often appear as small spots \cite{kou2023infrared}. Consequently, infrared small target detection (IRSTD) is essential for extracting valuable information from complex backgrounds.

\begin{figure}[htbp]
	\centering
	\includegraphics[width=1.0\linewidth]{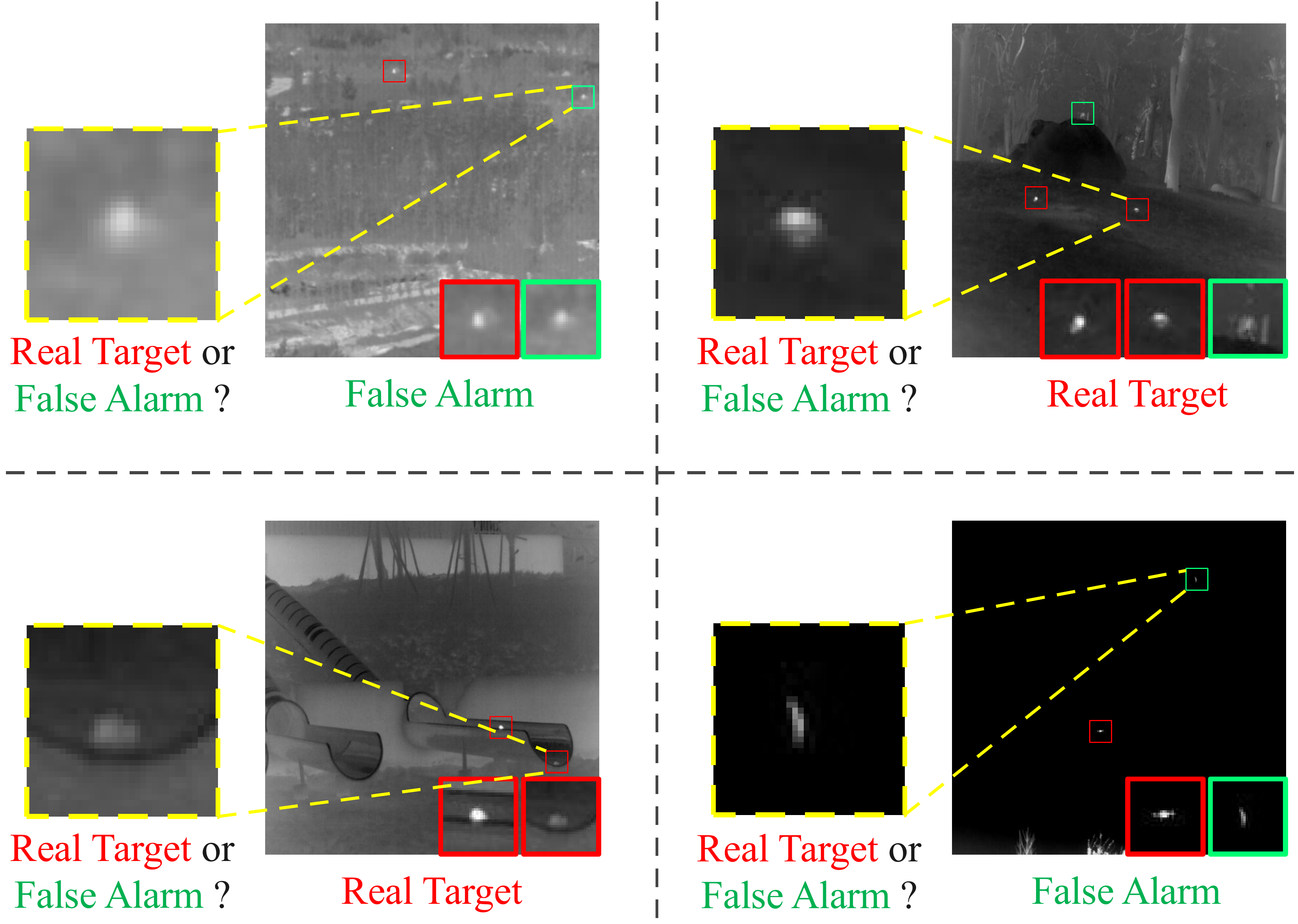}
	\caption{Due to the long imaging distance, real targets (red boxes) lack clear texture and structure, appearing as tiny bright spots. Meanwhile, high-brightness clutter in the background, such as from clouds, buildings, or the ground (green boxes), actively mimics real targets in local morphology, size, and brightness, creating indistinguishable interference.}
	\label{fig:fig1}
\end{figure}

Despite the critical role of IRSTD in infrared warning systems, as illustrated in Fig. \ref{fig:fig1}, two fundamental challenges remain in the feature extraction of infrared small targets. First, the low resolution of infrared imaging systems results in a paucity of texture and color information. Second, the inherent limitations of imaging small targets occupy an exceedingly small number of pixels. These challenges substantially impede practical performance in complex environments \cite{zhao2022single}. The primary issues impeding the detection of infrared small targets are as follows:
\begin{enumerate}
	\item \textbf{Insufficient features:} Infrared small targets often lack distinct structural features due to their low resolution and the absence of texture and color information, making it difficult to differentiate them from complex backgrounds.
	\item \textbf{Low signal-to-noise ratio:} The weak infrared radiation signal from small targets, combined with low contrast against background noise, increases the difficulty of detection and recognition.
\end{enumerate}
Addressing these challenges is essential for achieving robust and precise detection in infrared warning systems. To improve IRSTD performance, it is crucial to suppress background clutter. This can be achieved by enhancing the extraction of inherent target features and integrating contextual analysis.

Traditional data-driven IRSTD methodologies can be categorized into three primary approaches. They are Background Suppression-based (BS-based), Human Visual System-based (HVS-based), and Low Rank Matrix-based (LRM-based) methods. The BS-based methods \cite{tom1993morphology, zhu2020balanced, lu2022enhanced, deng2021entropy} utilize specialized filters to extract small targets. The HVS-based methods \cite{chen2013local, wei2016multiscale, qiu2022global} identify small targets by exploiting local information features. The LRM-based methods \cite{dai2017reweighted, zhang2018infrared} detect small targets through the decomposition and subsequent reconstruction of infrared images into sparse matrices. However, these traditional methods are limited in their ability to learn from datasets or dynamically adjust parameters in response to changing environmental conditions, leading to suboptimal performance in complex scenarios.

Recent advances in deep learning have yielded substantial improvements for data-driven IRSTD \cite{lin2024cs}. Previous research in deep learning methods emphasized weak feature expressions, such as gradient edge \cite{zhu2024towards}, contextual information \cite{dai2021asymmetric, wu2022uiu, li2022dense, lu2025infrared}, and local contrast \cite{dai2021attentional, zhang2023attention}. Despite their progress, fundamental challenges in IRSTD remain unaddressed. Specifically, the current limitations are primarily evident in three aspects:

\begin{enumerate}
	\item \textbf{Insufficient Feature Extraction}: Current convolution methods have shortcomings in capturing details of infrared small targets. Shallow networks have limited receptive fields, which struggle to capture non-local relationship between small targets and background. Meanwhile, deep networks may lose features due to downsampling, leading to inaccurate discrimination between targets and false alarms, thereby affecting the accuracy of detection.
	\item \textbf{Static Attention for Interference Suppression}: Traditional convolution and pooling strategies often struggle to effectively separate background and target features, while the saliency of infrared targets may be distributed in specific channels (such as thermal radiation intensity, gradient changes, etc.). However, existing attention mechanisms (such as SE \cite{bae2010small}, CBAM \cite{li2022dense}) model channel relationships statically or at a single scale \cite{lu2025infrared}, causing target information to be masked by background signals, reducing detection sensitivity.
	\item \textbf{Static Feature Fusion}: Current feature fusion methods, such as skip connected feature addition or concatenation \cite{ronneberger2015u}, ignore the differences in various target features at different levels, which can lead to shallow-level noise contamination of deep semantics features, and thus limit the network's adaptability in dynamic scenes. This further leads to a high false positive rate, affecting the reliability of the detection.
\end{enumerate}
These limitations collectively impede robust detection in real-world scenarios, where targets exhibit dynamic scales and unpredictable noise patterns.

Facing the challenges encountered in existing infrared small target detection, this study sets its core objective: to construct a high-performance detection framework capable of synergistically optimizing feature learning, contextual awareness, and feature fusion. Firstly, a dynamic feature extraction mechanism is introduced. This mechanism aims to resolve the inherent conflict between local receptive fields and global contextual modeling, adaptively adjusting its information aggregation scope to enhance the representational capacity of foundational features. Secondly, a multi-scale context-guided attention strategy is employed. This strategy first expands the perceptual range of local features and conducts intelligent grouping and refinement along the channel dimension based on the physical properties of infrared imaging, thereby generating multi-scale attention maps. Finally, to address missed detections and false alarms caused by improper fusion, an adaptive feature fusion method is proposed. This method dynamically evaluates and balances the contributions of features from different hierarchical levels, intelligently integrating complementary information. In summary, this research is committed to achieving breakthroughs in three critical aspects—feature extraction, information perception, and feature fusion—through the above synergistic approaches. The ultimate goal is to establish a robust infrared small target detection system capable of delivering high-precision and high robustness in complex, dynamic infrared scenarios.

To address the aforementioned challenges, we propose DCGANet, which consists of three core innovations.
First, to improve the recognition of real targets, suppress background noise, and discriminate between them and false alarms, we proposed \textbf{selective variable convolution (SVC)}, which consists of multi-branch adaptive kernels. The SVC resolves the local-global trade-off, which can accurately extract features of weak targets and significantly reduce background noise.
Second, we proposed a Dynamic Content-Guided Attention (DCGA) mechanism. This mechanism simulates the "coarse-to-fine" search pattern of the human visual system. Through a two-stage process, it dynamically generates a dedicated Spatial Importance Map (SIM) for each channel of the feature map, thereby precisely enhancing target features while suppressing background interference.
Last, to address the issue of improper fusion between different hierarchical features, we designed the Adaptive Dynamic Feature Fusion (ADFF) module. It replaces traditional skip connections, intelligently weighing and merging features from the encoder and decoder through a dynamic spatial selection mechanism. This effectively suppresses the contamination of deep semantic features by shallow-level noise.

The main contributions of this paper are listed as follows:
\begin{enumerate}
	\item We propose \textbf{DCGANet}, a novel framework that unifies feature learning, contextual guide attention, and feature fusion in a hierarchical dynamic pipeline. This framework bridges the gap between rigid CNN designs and the dynamic, sparse infrared target, achieving joint optimization of accuracy, robustness, and interpretability.
	\item The core of \textbf{DCGA} is to simulate the process of "rapid scanning to fine focusing" in HVS. By using the physical characteristics of infrared imaging, such as the differences in response to targets in different bands, potential target areas were prioritized and located through channel grouping refinement, while irrelevant backgrounds were suppressed.
	\item Our \textbf{SVC} combines deformable convolution, multi-dilated convolution, and standard convolution. This architecture enhances the discrimination between real targets and false alarms, and suppresses background noise through its broad receiving domain.
	\item The \textbf{ADFF} substantially enhances the main network's receptive field capacity to differentiate between actual targets and false positives via its sophisticated dynamic fusion technology.
\end{enumerate}

\section{Related Work} \label{sec:related}

\subsection{Infrared Small Target Detection}
The methodological evolution of infrared small target detection (IRSTD) is divided into model-driven and data-driven paradigms.

\subsubsection{Model-Driven Infrared Small Target Detection}

Model-driven methods formalize IRSTD as anomaly recognition against backgrounds, leveraging three computational paradigms: filters-based, human visual system (HVS) simulations-based, and sparse matrix reconstruction. The filter-based method \cite{tom1993morphology, zhu2020balanced, lu2022enhanced} mainly focuses on suppressing background clutter and highlighting targets while suppressing isolated noise through filter stacking. For instance, multi-scale Top-Hat \cite{deng2021entropy, deng2021infrared} filtering algorithm, which combines multi-scale transformation, can improve detection performance. However, when the infrared image contains complex backgrounds, the algorithm will exhibit a lower signal-to-noise ratio (SNR). To further improve the performance, the Human Visual System-Based (HVS-Based) methods \cite{chen2013local, wei2016multiscale, qiu2022global, qiu2020adaptive} utilize the principle of human visual contrast to suppress background brightness and enhance local information of infrared small targets, such as LCM \cite{chen2013local}, and TLLCM \cite{han2019local}. However, HVS-based methods still struggle to maintain detection stability when faced with high interference from bright edges or points. From another perspective, sparse matrix reconstruction treats small targets as sparse matrices and backgrounds as low-rank matrices \cite{dai2017reweighted, zhang2018infrared, zhang2019infrared}, detecting small targets by decomposition low-rank matrices. For instance, the Infrared Patch Image (IPI) \cite{gao2013infrared} model divides infrared images into blocks and recombines them into low-rank sparse matrices. Meanwhile, the reweighted infrared patch tensor (RIPT) \cite{dai2017reweighted} was proposed to construct low-rank sparse matrices more stably and faster through the alternating direction method. Overall, although these three categories of methods can achieve good detection results for certain images, their performance dramatically decreases when facing complex backgrounds.

\subsubsection{Data-Driven Infrared Small Target Detection}

Data-driven methods, empowered by deep learning advancements, have emerged as a research frontier in IRSTD. Data-driven methods improve original convolutional neural networks (CNNs) with the prior information of infrared small targets. On the one hand, for the enhancement of local information, the attention local contrast (ALCNet) Net \cite{dai2021attentional} was designed to enhance the local contrast of feature maps. Meanwhile, the dynamic attention transformer (DATNet) \cite{hu2025datransnet} was proposed to improve the performance through contextual connectivity. Furthermore, the gated-trans unet (GSTUnet) \cite{zhu2024towards} focuses on the local edge of the small target to improve detection performance. On the other hand, in optimizing network architectures for IRSTD, the dense nested attention (DNANet) \cite{li2022dense} employs U-shaped network topology with dense skip connections to achieve feature fusion and target enhancement. Additionally, the background-aware feature exchange net (BAFENet) \cite{xiao2024background} combines background semantics with feature exchange of targets and backgrounds to achieve high-precision detection.

Although deep learning methods have made substantial progress, they still face challenges such as over-parameterized architectures and inflexible receptive fields \cite{hu2025datransnet}, impending comprehensive capture of the details of small targets. Excessive reliance on local information \cite{zhang2022isnet, zhu2024tmp} also limits their ability to model global contextual features, leading to high false alarm rates in noisy scenarios. DCGANet distinguishes itself through two core innovations:

\begin{enumerate}
	\item \textbf{Multi-convolutional spatiotemporal feature extraction:} Overcoming CNN rigidity via adaptive kernel configurations across spectral-temporal domains.

	\item \textbf{HVS-mimetic dynamic attention:} The DCGA module implements 'coarse-to-fine' saliency mapping, mirroring human visual system search patterns through channel-specific attention refinement.
\end{enumerate}

\subsection{Vision Attention Mechanisms on IRSTD}

Vision attention mechanism in IRSTD enhance target feature expression representation by dynamically prioritizing regions, simultaneously amplifying discriminative target and effectively suppressing noise and false alarms. For instance, the local patch network with global attention net (LPNet) \cite{chen2022local} improves the detection accuracy of small targets by combining global and local features. The infrared shape network (ISNet) \cite{zhang2022isnet} used Taylor finite difference heuristic blocks and bidirectional attention aggregation modules. In addition, the dynamic gradient attention transformer net (DATransNet) \cite{hu2025datransnet} further improves detection performance by introducing a transformer detection module and gradient attention.

Despite these advancements, existing methods still lack robust global feature understanding in IRSTD. In contrast, Dynamic Content-Guided Attention (DCGA) implements a coarse-to-fine attention cascade: it first generates a rough spatial attention map, then refines it into a fine-grained map by integrating spatial and channel attention. This process refines initial spatial attention maps (SIM) for each channel to focus on the most representative features, enhancing the recognition of suspected small targets. In feature fusion, DCGA adopts a shallow-deep feature fusion strategy to improve information flow between network layers, preserving features and enabling effective gradient backpropagation. This dual optimization not only boosts small target detection accuracy but also strengthens false alarm suppression capabilities.

\section{Methodology}

\label{sec:method}

\begin{figure*}[htbp]
	\centering
	\includegraphics[width=1.0\linewidth]{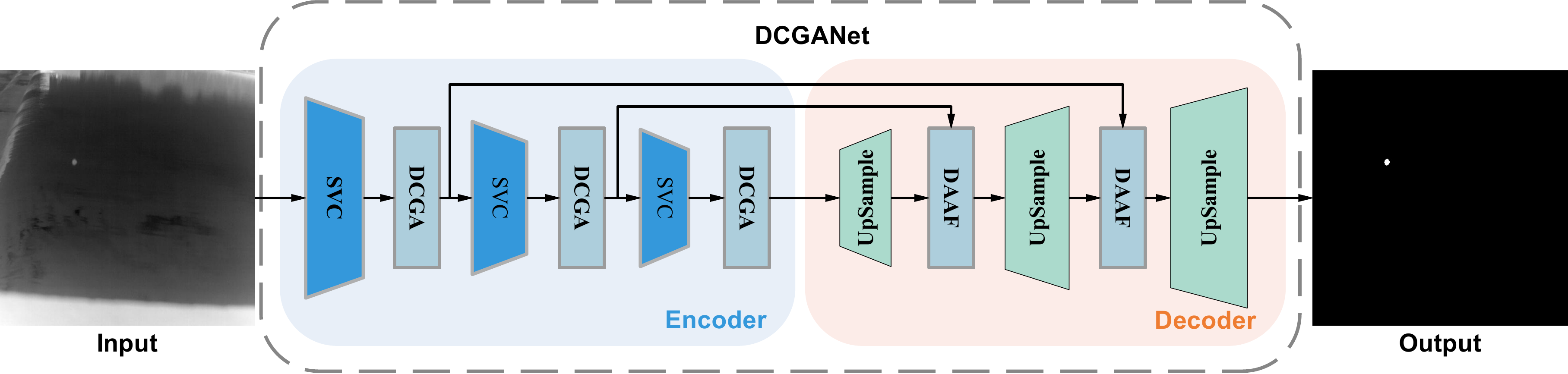}
	\caption{The overall architecture of the proposed DCGANet. It is built upon a U-Net-like backbone and integrates the key Selective Variable Convolution (SVC), Dynamic Context-Guided Attention (DCGA), and Adaptive Dynamic Feature Fusion (ADFF) modules.}
	\label{fig:DCGANet}
\end{figure*}

\begin{figure}[htbp]
	\centering
	\includegraphics[width=1.0\linewidth]{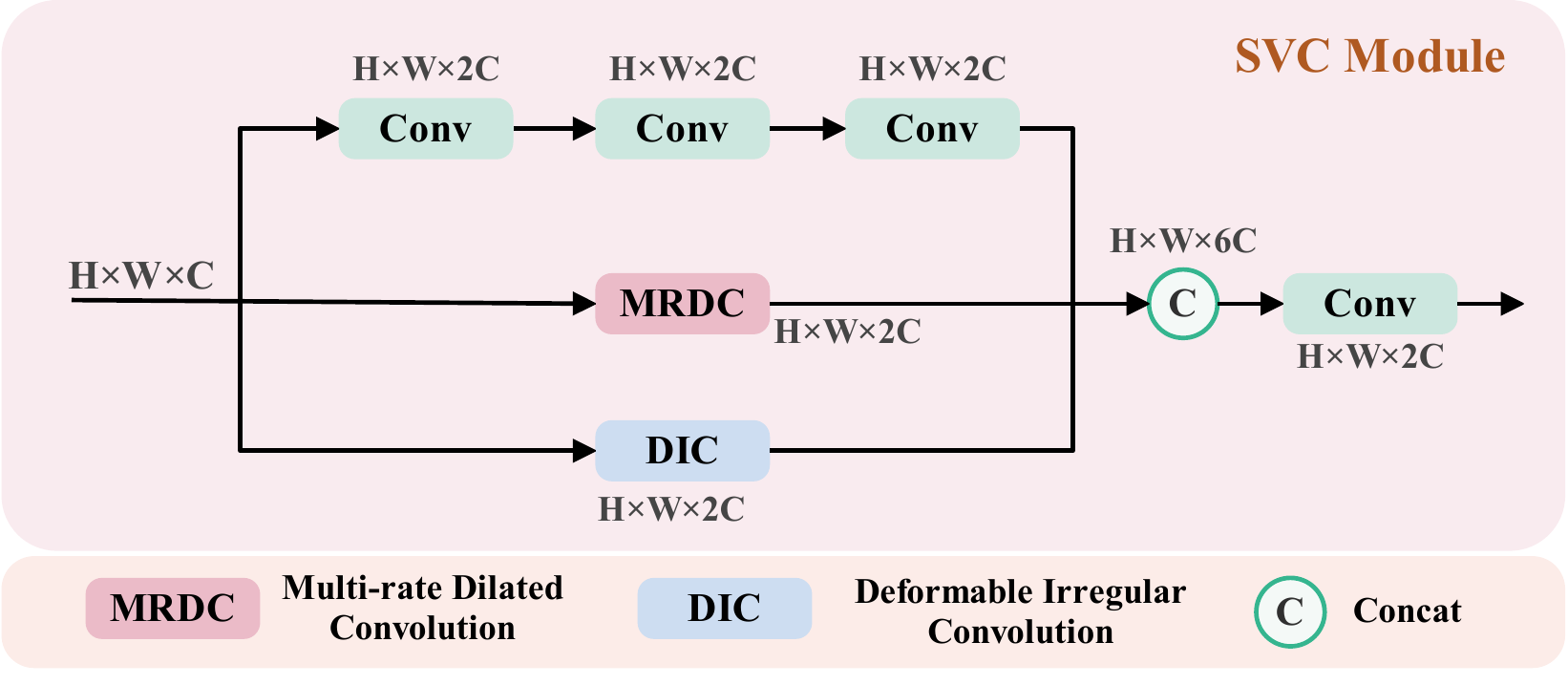}
	\caption{The structure of the SVC module: a parallel branch architecture that performs different types of convolution operations on the input feature map. This design enables the encoding stage to extract features at different scales and preserve details.}
	\label{fig:SVC}
\end{figure}

The overall architecture of DCGANet is illustrated in Figure~\ref{fig:DCGANet}, built upon a U-Net-like framework and composed of three core components: the Selective Variable Convolution (SVC), Dynamic Context-Guided Attention (DCGA), and Adaptive Dynamic Feature Fusion (ADFF) modules. Specifically, the encoder path consists of three cascaded layers, each sequentially integrating an SVC and a DCGA module. SVC replaces standard convolution to enhance faint target features while expanding the receptive field for stronger contextual modeling. The subsequent DCGA mechanism generates channel-specific Spatial Importance Maps (SIMs) through input-guided attention calibration, enabling per-channel saliency emphasis crucial for sub-pixel infrared targets. On the decoder path, two ADFF modules dynamically fuse multi-scale features from the encoder, replacing the original skip connections and effectively suppressing noise through a context-aware fusion strategy.

\subsection{Selective Variable Convolution}

In Convolutional Neural Networks (CNNs), the limited receptive field of traditional convolutional layers hinders their ability to capture global contextual information in Infrared Small Target Detection (IRSTD). To address this, we propose the Selective Variable Convolution (SVC) module, a powerful foundational block designed to significantly enhance the network's feature learning capabilities.

The structure of SVC, shown in Figure~\ref{fig:SVC}, enhances expressive power by integrating features from three parallel branches.
\begin{itemize}
	\item \textbf{Standard Convolution (SConv):} The first branch employs a standard convolution to preserve local spatial features, maintaining the stability of basic patterns for small targets and foundational detection performance.
	\item \textbf{Deformable Convolution (DConv):} The second branch utilizes deformable convolution, which excels at outlining the edges of irregularly shaped targets, enhancing target features at different scales.
	\item \textbf{Multiple Dilation Convolution (MDConv):} The third branch is a multiple dilation convolution, which uses kernels with different dilation rates (rates = 2, 4, 8) to expand the receptive field, thereby aggregating multi-scale context to suppress structured clutter.
\end{itemize}
The mathematical expressions for these three branches are as follows:
\begin{align}
	X_s & = \operatorname{SConv}(X_{\text{in}}),  \\
	X_c & = \operatorname{DConv}(X_{\text{in}}),  \\
	X_m & = \operatorname{MDConv}(X_{\text{in}}),
\end{align}
where $X_{\text{in}}$ is the input feature, and $X_s$, $X_c$, and $X_m$ are the outputs of the three branches, respectively.

The outputs of these branches are fused via channel-wise concatenation and a $1 \times 1$ convolution, yielding an integrated representation enriched in both local precision and global differentiability.
\begin{equation}
	X_{\text{out}} = \operatorname{Conv}_{1 \times 1} \left( \operatorname{Concat} \left( X_s, X_c, X_m \right) \right),
\end{equation}
where $X_{\text{out}}$ is the final output of the SVC module. Through this multi-convolution strategy, SVC effectively improves the recognition of target edges and details, and its expanded receptive field and deformable feature extraction provide significant advantages for context-aware modeling and noise suppression.

\subsection{Dynamic Context-Guided Attention}

Although the SVC module enhances feature extraction, it may still face difficulties when dealing with complex background clutter. Traditional attention mechanisms suffer from two key problems: first, a lack of information interaction between channel and spatial attention; and second, a single spatial weight for all channels cannot adapt to the diverse distributions across multiple feature channels, risking the dilution of key target information.

\begin{figure}[htbp]
    \centering
    \includegraphics[width=\linewidth]{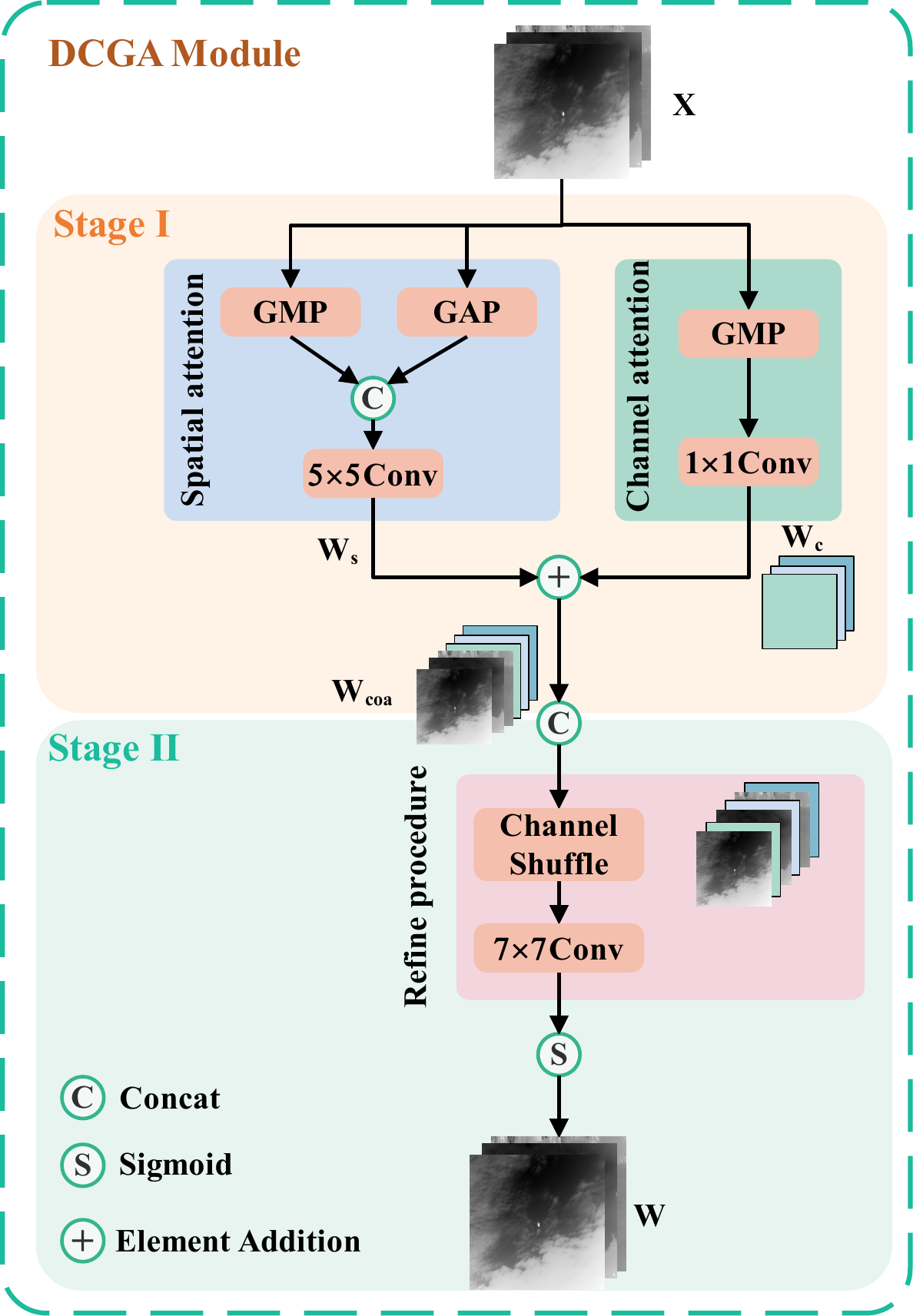}
    \caption{The structure of the DCGA module: It employs a parallel approach to equally weight spatial and channel information. It also incorporates a channel shuffling of channel and spatial information to enhance target segmentation capabilities.}
    \label{fig:CGA}
\end{figure}

To address these challenges, we propose the Dynamic Context-Guided Attention (DCGA) module. The structure of DCGA, shown in Figure~\ref{fig:CGA}, employs a coarse-to-fine approach to Spatial Importance Map (SIM), effectively simulating the Human Visual System's (HVS) ability to focus on salient features and suppress false positives.

\subsubsection{Stage 1: Coarse-Grained Attention and Joint Channel-Spatial Fusion}
This stage generates a preliminary attention map.

\textbf{Channel Attention:} A dual-bottleneck structure is used, aggregating spatial information via Global Average Pooling (GAP) to extract the global channel response $W_c$:
\begin{equation}
	W_c = \operatorname{\mathcal{C}}_{1 \times 1} \left( \operatorname{ReLU} \left( \operatorname{\mathcal{C}}_{1 \times 1} \left( X_{GAP}^{\text{c}} \right) \right) \right),
\end{equation}
where $X_{GAP}^{\text{c}}$ is the channel-compressed feature obtained from Global Average Pooling.

\textbf{Spatial Attention:} Spatial saliency $W_s$ is captured by concatenating multi-view features from Global Average Pooling (GAP) and Global Max Pooling (GMP):
\begin{equation}
	W_s = \operatorname{\mathcal{C}}_{7 \times 7} \left( \operatorname{Concat} \left( X_{GAP}^{\text{s}}, X_{GMP}^{\text{s}} \right) \right),
\end{equation}
where $X_{GAP}^{\text{s}}$ and $X_{GMP}^{\text{s}}$ represent the spatial descriptors from GAP and GMP, respectively.

\textbf{Interactive Attention:} Following broadcasting rules, we fuse $W_c$ and $W_s$ using element-wise addition to obtain a coarse SIM $W_{\text{coa}}$.
\begin{equation}
	W_{\text{coa}} = W_c \oplus W_s,
\end{equation}
where $W_{\text{coa}} \in \mathbb{R}^{C \times H \times W}$, and $\oplus$ denotes element-wise addition with automatic dimension expansion.

\subsubsection{Stage 2: Fine-Grained Calibration for Dynamic SIM Generation}
This stage refines the coarse attention map.

\textbf{Channel-Shuffle Fusion:} The original features $X \in \mathbb{R}^{C \times H \times W}$ and the coarse attention map $W_{coa}$ are concatenated and then shuffled along the channel dimension\cite{zhang2018shufflenet}, thereby establishing cross-channel correlations while preserving spatial details.

\textbf{Dynamic Weight Generation:} A gated convolutional layer $\mathcal{GC}_{7 \times 7}$ followed by a Sigmoid activation function generates the final channel-specific SIM $W$:
\begin{equation}
	W = \sigma \left( \operatorname{\mathcal{GC}}_{7 \times 7} \left( \operatorname{CS} \left( \operatorname{Concat} \left( X, W_{\text{coa}} \right) \right) \right) \right),
\end{equation}
where $\operatorname{CS}(\cdot)$ denotes the channel shuffle operation and $\sigma(\cdot)$ is the Sigmoid function. This content-adaptive calibration allows the SIM for each channel to be dynamically adjusted based on local feature patterns, guiding the model to focus on important regions within each channel.

\begin{figure}
    \centering
    \includegraphics[width=0.6\linewidth]{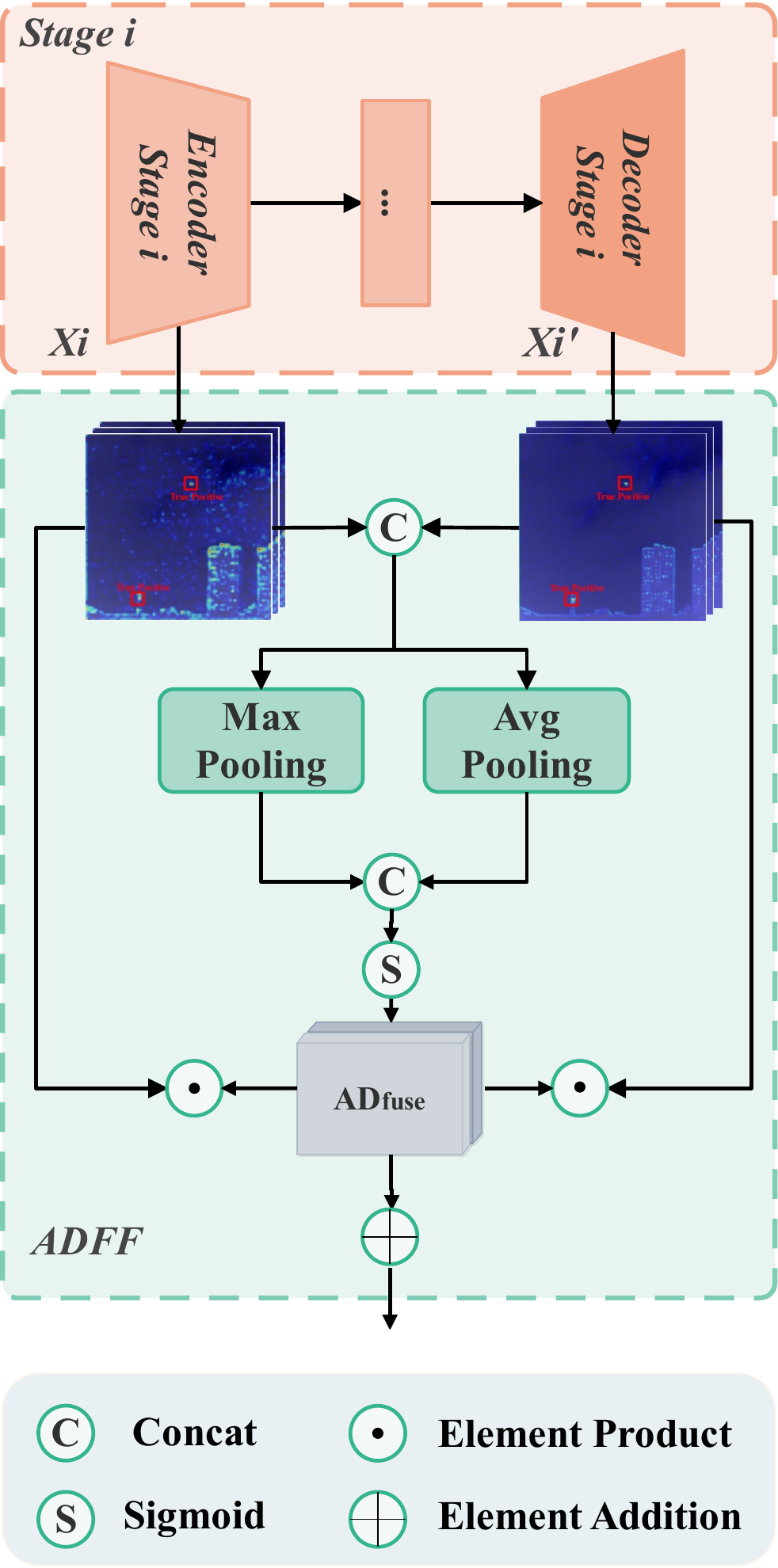}
    \caption{The structure of the ADFF module: It integrates the encoding and decoding stages of the network via a spatially adaptive selection mechanism. It utilizes cross-channel max-pooling and average-pooling for precise target extraction.}
    \label{fig:ADFFM}
\end{figure}

\subsection{Adaptive Dynamic Feature Fusion}

Our proposed Adaptive Dynamic Feature Fusion (ADFF) module is designed to enhance the network's ability to distinguish real targets from similar false positives. As shown in Figure~\ref{fig:ADFFM}, we introduce a dynamic spatial selection mechanism into the original skip connection, leveraging multi-scale features to enable the network to focus on regions richest in spatial context. The ADFF module is specifically designed to process the encoder features $X_i$ and decoder features $X'_i$ at stage $i$:
\begin{equation}
	X_i, X'_i \in \mathbb{R}^{C_i \times \frac{H}{2^i} \times \frac{W}{2^i}},
\end{equation}
where $C_i$ is the number of channels, and $\frac{H}{2^i}$ and $\frac{W}{2^i}$ are the height and width of the feature maps.

We first apply channel-wise average-pooling and max-pooling operations to the concatenated feature map:
\begin{align}
	AD_i^{\text{avg}} & = \operatorname{AvgPool} \left( \operatorname{Concat} \left[ X_i, X'_i \right] \right), \\
	AD_i^{\text{max}} & = \operatorname{MaxPool} \left( \operatorname{Concat} \left[ X_i, X'_i \right] \right),
\end{align}
where $\operatorname{AvgPool}$ and $\operatorname{MaxPool}$ are the average-pooling and max-pooling operations, respectively.

Next, we concatenate these two pooled features and pass them through a convolutional layer and a Sigmoid function to generate the final spatial attention map $AD_{\text{fuse}}$:
\begin{equation}
	AD_{\text{fuse}} = \sigma \left( \operatorname{Conv}_{k \times k} \left( \operatorname{Concat} \left( AD_i^{\text{avg}}, AD_i^{\text{max}} \right) \right) \right),
\end{equation}
where $\operatorname{Conv}_{k \times k}$ is a convolutional layer with a kernel size of $k \times k$ (e.g., $7 \times 7$).

The final output $Y_i$ of the ADFF module is obtained by element-wise multiplication of the generated attention map $AD_{\text{fuse}}$ with the input features $X_i$ and $X'_i$ respectively, and then adding the results:
\begin{equation}
	Y_i = (X_i \odot AD_{\text{fuse}}) + (X'_i \odot AD_{\text{fuse}}),
\end{equation}
where $\odot$ denotes element-wise multiplication. Through this adaptive selective fusion strategy, the ADFF module can automatically identify and integrate the most discriminative features, thereby enhancing the network's representation capability in infrared small target detection.

\begin{table*}[b]
	\centering
	\footnotesize
	\caption{Configurations for all comparative experiments}
	\label{tab:parameters_config} % Renamed from tab:parameters to be unique
	\resizebox{\textwidth}{!}{
		\begin{tabular}{ll}
			\toprule[1.5pt]
			\textbf{Methods (\textit{Source \& Year})}          & \textbf{Key parameters configurations}                                                                       \\
			\midrule
			\multicolumn{2}{l}{\textit{Background Suppression Methods}}                                                                                                        \\
			\midrule
			NWMTH \cite{bai2010analysis} \textit{(PR'2010)}                           & $R_{o}=9,R_{i}=4$                                                                                            \\
			FKRW \cite{qin2019infrared} \textit{(TGRS'2019)}                           & Windows size: 11, $K=4,p=6,\beta=200$                                                                        \\
			\midrule
			\multicolumn{2}{l}{\textit{HVS-based Methods}}                                                                                                                     \\
			\midrule
			MPCM \cite{wei2016multiscale}\textit{(PR'2016)}                             & $N = 1, 3, \cdots, 9$, threshold $k = 3$                                                                     \\
			WLDM \cite{deng2016small}\textit{(TGRS'2016)}                           & $L = 4, m = 2, n = 2$, threshold $k = 2$                                                                     \\
			RLCM \cite{han2018infrared}\textit{(GRSL'2018)}                           & $k_1 = [2,5,9], k_2 = [4, 9, 16]$, scale: 3, threshold $k = 1$                                               \\
			ILCM \cite{han2020infrared}\textit{(GRSL'2020)}                           & Cell size: $3 \times 3$, threshold $k = 3$                                                                   \\
			TLLCM \cite{han2019local}\textit{(GRSL'2020)}                          & Scales:[5,7,9] $\lambda =0.5$, threshold $ k=3$                                                              \\
			GSWLCM \cite{qiu2022global}\textit{(GRSL'2022)}                         & Local Window Structures: $[3,5,7,9]$, $\delta=0.01, k = 20$                                                  \\
			\midrule
			\multicolumn{2}{l}{\textit{Low-rank and Sparse Decomposition Methods}}                                                                                             \\
			\midrule
			IPI \cite{gao2013infrared}\textit{(TIP'2013)}                             & Patch size: 50, sliding step: 10, $\lambda=1/\sqrt{\mathrm{min}(m,n)}$, $\epsilon=10^{-7}$                   \\
			RIPT \cite{dai2017reweighted} \textit{(JSTARS'2017)}                         & Patch size: 50, sliding step: 10, $\lambda=2/\sqrt{\mathrm{min}(m,n)}$, $\epsilon=10^{-2}$, $\omega=10^{-7}$ \\
			PSTNN  \cite{zhang2019infrared}\textit{(RS'2019)}                            & Patch size: 40, sliding step: 40, $\lambda=0.6/\sqrt{\mathrm{max}(n_1,n_2)\times n_3}$, $\epsilon=10^{-7}$   \\
			\midrule
			\multicolumn{2}{l}{\textit{Deep Learning Methods}}                                                                                                                 \\
			\midrule
			ACM  \cite{dai2021asymmetric}\textit{(WACV'2021)}                            & Backbone: FPN, layer blocks: [4, 4, 4], channels: [8, 16, 32, 64]                                            \\
			AGPCNet \cite{zhang2023attention}\textit{(TAES'2023)}                        & Backbone: resnet34, scales: [10, 6, 5, 3], reduce ratios: [16, 4], gca type: patch, gca att: post            \\
			DNANet  \cite{li2022dense}\textit{(TIP'2022)}                          & Backbone: resnet18, layer blocks: [2, 2, 2, 2], filter: [16, 32, 64, 128, 256]                               \\
			ISNet \cite{zhang2022isnet}\textit{(CVPR'2022)}                          & Backbone: resnet18, layer blocks: [4, 4, 4], channels: [8, 16, 32, 64], fuse mode: AsymBi                    \\
			UIUNet \cite{wu2022uiu}\textit{(TIP'2023)}                          & Channels: [64, 128, 256, 512], fuse mode: AsymBi                                                             \\
			MTUNet \cite{wu2023mtu}\textit{(TGRS'2023)}                         & Backbone: resnet18, layer blocks: [2, 2, 2, 2], filter: [16, 32, 64, 128, 256]                               \\
			SeRankDet \cite{dai2024pick}\textit{(TGRS'2024)}                      & Channels:[64, 128, 256, 512, 1024],o=3                                                                       \\
			SCTransNet \cite{yuan2024sctransnet}\textit{(TGRS'2024)}                     & Backbone:Swin-Tiny,  Layer Num=[2,6,6,2], Head Num=[3,6,12,24]                                               \\
			\rowcolor[gray]{0.9}$\star$ \textbf{DCGANet (Ours)} & Channels: [64, 128, 256, 512, 1024]                                                                          \\
			\bottomrule[1.5pt]
		\end{tabular}}
\end{table*}

\section{Experiments}

\subsection{Experimental Setup}

\subsubsection{Datasets}
To rigorously evaluate the performance and generalization capabilities of our method, we conduct comprehensive benchmark experiments on three widely recognized infrared small target detection datasets: NUAA-SIRST, IRSTD-1K, and NUDT-SIRST. These datasets exhibit considerable diversity in target characteristics, background clutter, and imaging conditions, thereby enabling robust validation of detection methodologies.

\begin{itemize}
	\item \textbf{NUAA-SIRST}: This dataset contains 427 real-world images capturing diverse scenes, such as clouds, cities, and oceans. With 480 target instances, it serves as a foundational benchmark for evaluating detection performance in common scenarios.

	\item \textbf{IRSTD-1K}: Comprising 1,001 highly diverse images, IRSTD-1K is designed to challenge a model's cross-domain generalization. The complexity of its scene types and target shapes imposes significantly higher demands compared to other datasets.

	\item \textbf{NUDT-SIRST}: This synthetic dataset includes 1,327 images with complex backgrounds. By employing adaptive functions for target size and intensity blurring, it generates a wide variety of targets (point-like, patchy, extended), creating a distribution that closely mimics real-world data and tests model robustness against complex interference.
\end{itemize}

\subsubsection{Evaluation Metrics}
To quantitatively analyze our DCGANet and the comparative methods, we adopt four standard metrics: Intersection over Union (IoU), normalized IoU (nIoU), Probability of Detection ($P_d$), and False-alarm Rate ($F_a$).

\begin{itemize}
	\item \textbf{IoU} measures the localization accuracy by comparing the overlap between the predicted bounding box and the ground truth. It is defined as the ratio of their intersection to their union.
	      \begin{equation}
		      \text{IoU} = \frac{TP}{T + P - TP}
	      \end{equation}

	\item \textbf{nIoU} is a normalized adaptation of the IoU metric, averaging the IoU over all target-containing images to provide a dataset-level performance score.
	      \begin{equation}
		      \text{nIoU} = \frac{1}{N} \sum_{i=1}^{N} \frac{TP(i)}{T(i) + P(i) - TP(i)}
	      \end{equation}

	\item \textbf{$P_d$} (also known as Recall) quantifies the model's ability to correctly identify true targets. A higher value is better.
	      \begin{equation}
		      P_d = \frac{TP}{TP + FN}
	      \end{equation}

	\item \textbf{$F_a$} measures the rate of pixels incorrectly identified as targets. It is calculated per image, and a lower value is better.
	      \begin{equation}
		      F_a = \frac{FP}{\text{Number of Pixels in Image}}
	      \end{equation}
\end{itemize}
where $N$ is the total number of images with targets, $T$ and $P$ are the sets of ground-truth and predicted positive pixels, respectively. $TP$, $FP$, and $FN$ denote the counts of true positive, false positive, and false negative pixels. In addition, the Receiver Operating Characteristic (ROC) curve illustrates the trade-off between $P_d$ and $F_a$.

\subsubsection{Implementation Details}
We adopt U-Net as the baseline, Soft-IoU Loss as the loss function, and AdamW as the optimizer, incorporating a polynomial learning rate decay strategy. A deep supervision strategy is also employed to enhance performance. All experiments are conducted using distributed training on a system with four NVIDIA RTX 3090 GPUs. To ensure a fair and comprehensive evaluation, we train independent models for each of the three datasets. The hyperparameter configurations for all comparative methods are summarized in Table~\ref{tab:parameters_config}. Due to significant differences in data distribution and resolution, we customize the training hyperparameters for each dataset, as detailed in Table~\ref{tab:hyperparameters}.

\begin{table}[htbp]
	\centering
	\small
	\caption{Customized Training Hyperparameters for Each Dataset.}
	\label{tab:hyperparameters}
    \resizebox{\columnwidth}{!}{
	\begin{tabular}{lcccc}
		\toprule[1.5pt]
		\textbf{Dataset} & \textbf{Epochs} & \textbf{Learning Rate (Lr)} & \textbf{Batch Size} & \textbf{Resolution} \\
		\midrule
		NUAA-SIRST       & 500             & 1e-4                        & 4                   & $512 \times 512$    \\
		IRSTD-1K         & 500             & 1e-4                        & 4                   & $512 \times 512$    \\
		NUDT-SIRST       & 500             & 1e-4                        & 16                  & $256 \times 256$    \\
		\bottomrule[1.5pt]
	\end{tabular}}
\end{table}

\subsection{Comparison with State-of-the-Art Methods}
\label{sec:comparison}
We rigorously benchmark our DCGANet against a wide range of state-of-the-art (SOTA) methods, including traditional model-driven techniques (Background Suppression, HVS-based, and Low-rank Matrix-based) and various advanced deep learning models. The comprehensive results on all three datasets are summarized in Table~\ref{tab:sota_comparison}, leading to several key observations:

\begin{enumerate}
	\item \textbf{Superior Overall Performance}: Our proposed DCGANet consistently achieves the best or second-best performance across all metrics on all three datasets. This demonstrates its superior accuracy and robustness in handling the challenges of low signal-to-noise ratio (SNR) and indistinct target boundaries. The combination of its dynamic feature extraction and fusion modules proves highly effective.

	\item \textbf{Data-Driven vs. Model-Driven}: A significant performance gap (often >15\% in IoU) exists between deep learning (DL) methods and traditional approaches. This highlights the inherent limitations of model-driven techniques, which rely on fixed priors like local contrast or global sparsity. In contrast, data-driven DL methods show superior generalization by learning deep hierarchical representations, making them more adaptable and less sensitive to hyperparameters in complex, unstructured scenes.

	\item \textbf{Advantage over SOTA DL Methods}: On the challenging NUAA-SIRST and IRSTD-1K datasets, DCGANet surpasses recent powerful models like SeRankDet. For instance, on NUAA-SIRST, DCGANet achieves the highest IoU, nIoU, and $P_d$, with only a marginal trade-off in the $F_a$ rate compared to SeRankDet. This indicates that while other models may be effective, DCGANet's architecture provides a better balance of high detection probability and precise pixel-level localization, especially in cluttered environments.
\end{enumerate}

\begin{figure*}[htbp]
    \centering

    \subfloat[NUAA-SIRST\label{fig:roc_nuaa}]
    {
        \includegraphics[width=0.32\textwidth]{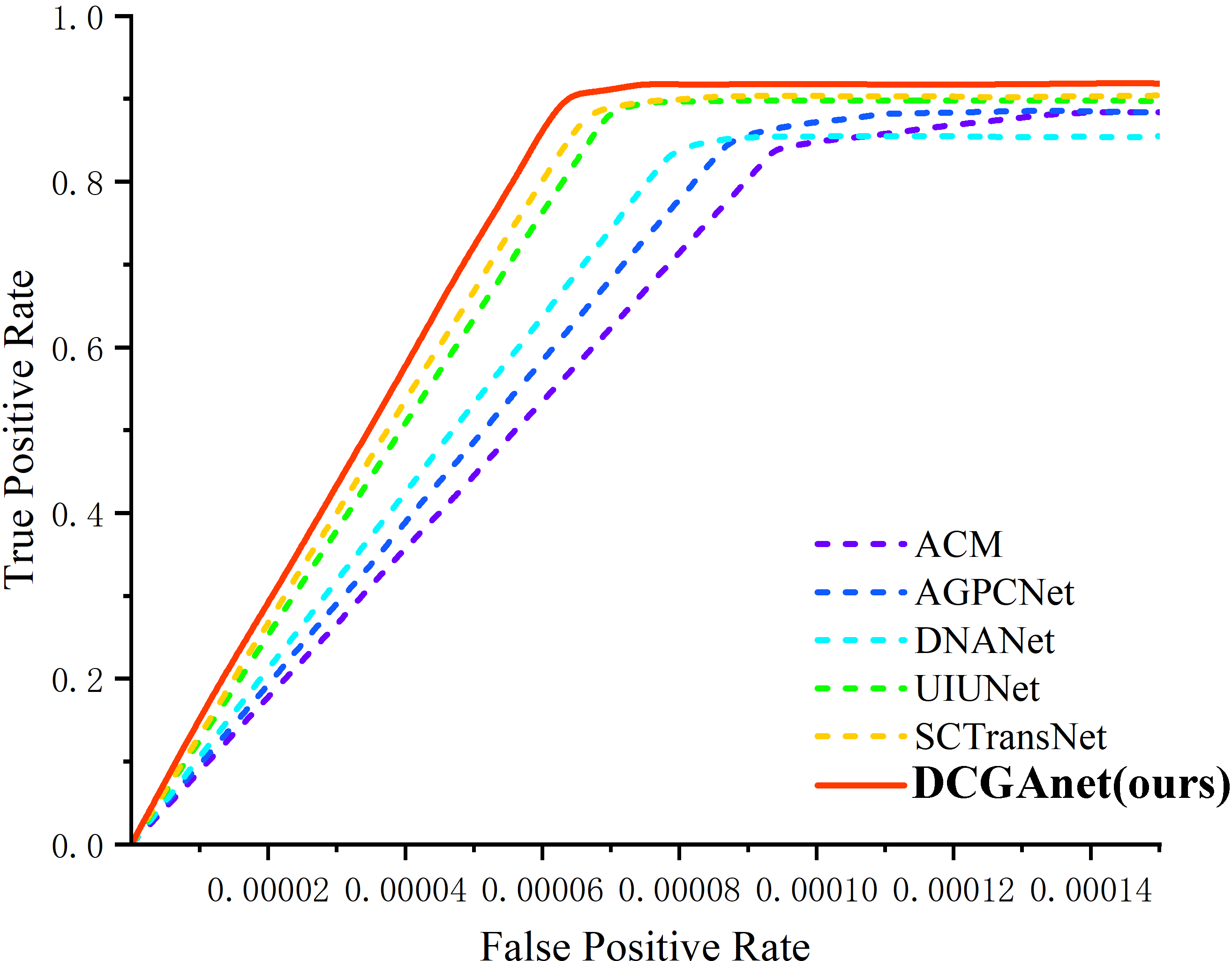}
    }
    \subfloat[IRSTD-1K\label{fig:roc_irstd}]
    {
        \includegraphics[width=0.32\textwidth]{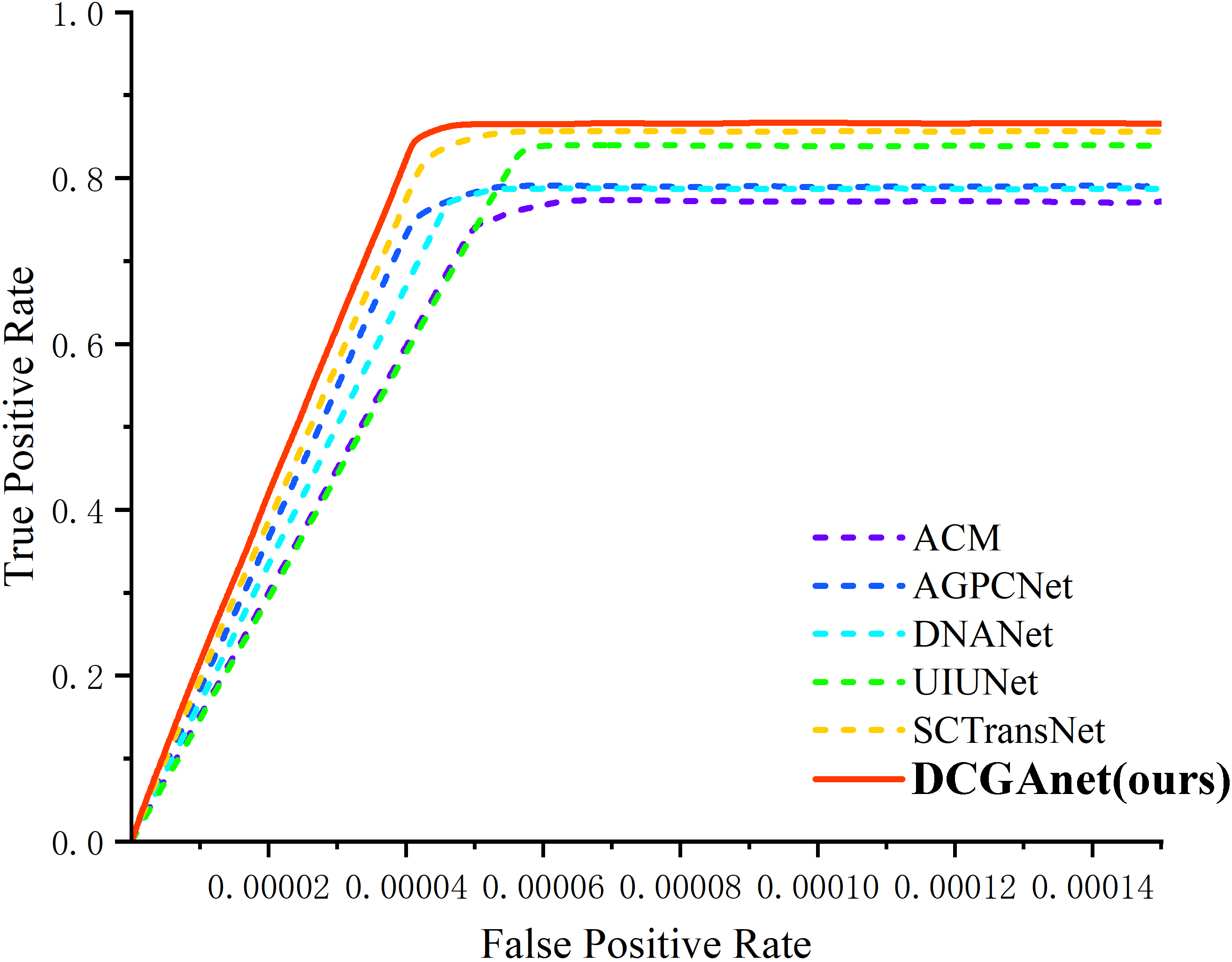}
    }
    \subfloat[NUDT-SIRST\label{fig:roc_nudt}]
    {
        \includegraphics[width=0.32\textwidth]{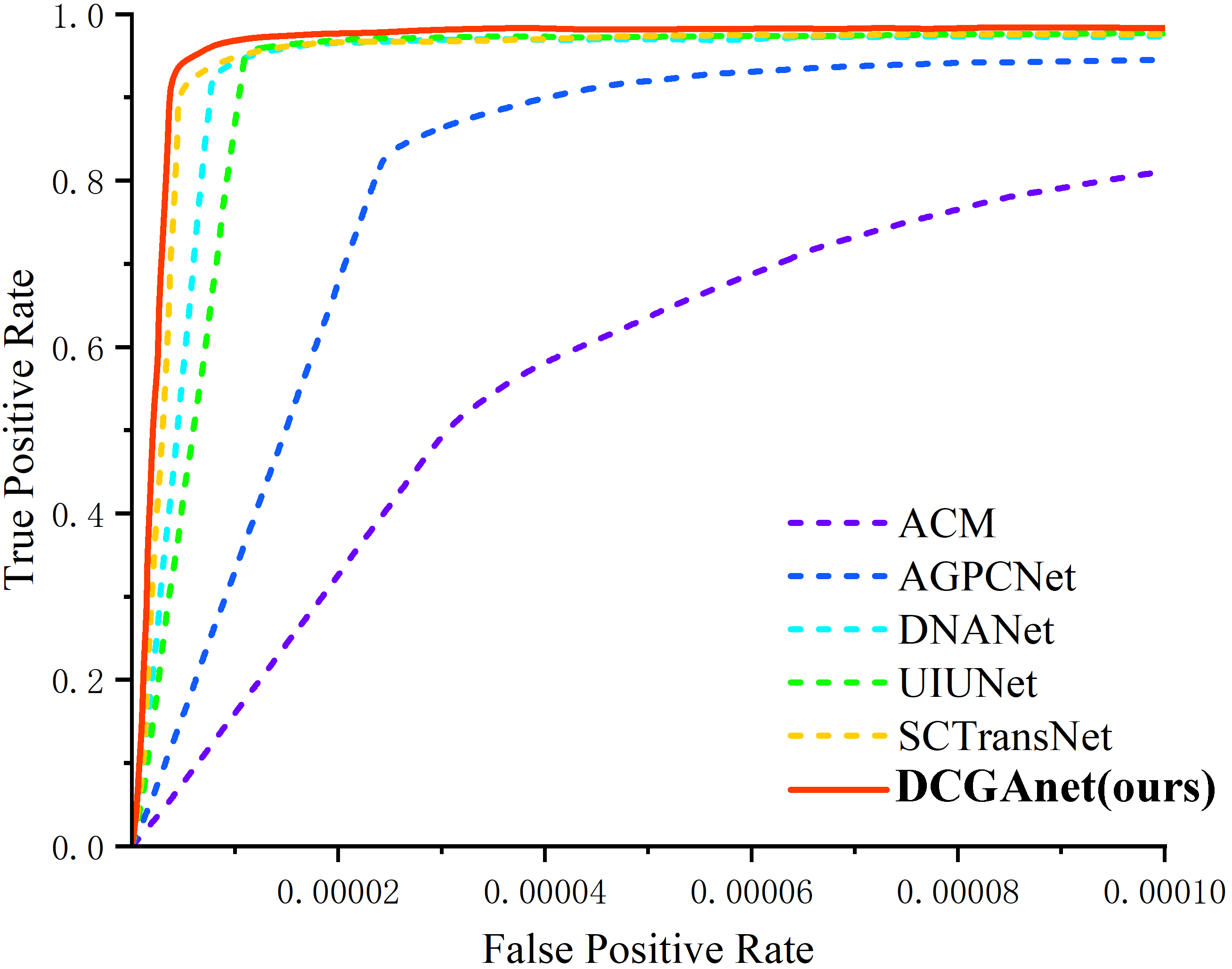}
    }

    \caption{ROC curves of different methods on the three datasets. Our DCGANet consistently achieves the highest $P_d$ at very low $F_a$ levels, demonstrating its superior trade-off between detection rate and false alarms.}
    \label{fig:roc}
\end{figure*}

Furthermore, Figure~\ref{fig:roc} illustrates the Receiver Operating Characteristic (ROC) curves for the three datasets. The analysis visually corroborates the quantitative results in Table~\ref{tab:sota_comparison}, confirming that DCGANet consistently achieves a higher probability of detection ($P_d$) at any given false-alarm rate ($F_a$), underscoring its stability and superiority.

\begin{table*}[htbp]
	\centering
	\scriptsize
	\caption{Quantitative comparison with state-of-the-art methods on three datasets. For all metrics except $F_a$, higher is better ($\uparrow$). For $F_a$, lower is better ($\downarrow$). The best performance is highlighted in \textcolor{red}{red}, and the second-best is in \textcolor{blue}{blue}.}
	\label{tab:sota_comparison}
	\setlength{\tabcolsep}{0.25cm} % Adjust column separation
	\resizebox{\textwidth}{!}{
		\begin{tabular}{l|cccc|cccc|cccc}
			\toprule[1.5pt]
			\multirow{2}{*}{\textbf{Method}}                     & \multicolumn{4}{c|}{\textbf{NUAA-SIRST} (Tr=60\%)} & \multicolumn{4}{c|}{\textbf{IRSTD-1K} (Tr=60\%)} & \multicolumn{4}{c}{\textbf{NUDT-SIRST} (Tr=60\%)}                                                                                                                                                                                                                                          \\
			                                                     & IoU$\uparrow$                                      & nIoU$\uparrow$                                   & $P_d$$\uparrow$                                   & $F_a$$\downarrow$      & IoU$\uparrow$           & nIoU$\uparrow$          & $P_d$$\uparrow$         & $F_a$$\downarrow$       & IoU$\uparrow$           & nIoU$\uparrow$          & $P_d$$\uparrow$         & $F_a$$\downarrow$       \\
					\midrule
			\multicolumn{13}{l}{\textit{Background Suppression Methods}}                                                                                                                                                                                                                                                                                                                                                                                              \\
					\midrule
			NWMTH  \cite{bai2010analysis}                                              & 14.74                                              & 16.33                                            & 68.29                                             & 55.18                  & 17.58                   & 16.08                   & 48.00                   & 21.05                   & 18.12                   & 15.58                   & 60.35                   & 32.44                   \\
			FKRW   \cite{qin2019infrared}                                              & 21.23                                              & 27.67                                            & 78.18                                             & 16.66                  & 9.72                    & 16.16                   & 67.11                   & 24.18                   & 9.64                    & 16.09                   & 69.44                   & 66.88                   \\
					\midrule
			\multicolumn{13}{l}{\textit{HVS-based Methods}}                                                                                                                                                                                                                                                                                                                                                                                                           \\
					\midrule
			MPCM    \cite{wei2016multiscale}                                              & 24.63                                              & 26.54                                            & 65.32                                             & 45.05                  & 19.90                   & 22.33                   & 56.45                   & 31.09                   & 26.31                   & 37.41                   & 53.34                   & 11.82                   \\
			WLDM   \cite{deng2016small}                                               & 5.33                                               & 3.23                                             & 62.40                                             & 21.21                  & 7.84                    & 9.91                    & 52.68                   & 14.79                   & 7.87                    & 11.17                   & 15.65                   & 64.11                   \\
			RLCM   \cite{han2018infrared}                                               & 24.55                                              & 26.37                                            & 82.92                                             & 62.13                  & 15.81                   & 20.96                   & 68.05                   & 67.67                   & 7.89                    & 10.97                   & 67.44                   & 31.65                   \\
			ILCM   \cite{han2020infrared}                                                & 17.69                                              & 18.40                                            & 36.50                                             & 40.61                  & 13.13                   & 10.89                   & 41.60                   & 47.65                   & 4.96                    & 6.22                    & 25.23                   & 88.20                   \\
			TLLCM   \cite{han2019local}                                               & 17.48                                              & 26.53                                            & 78.98                                             & 16.17                  & 10.43                   & 17.30                   & 66.22                   & 24.98                   & 12.91                   & 16.31                   & 73.97                   & 69.63                   \\
			GSWLCM     \cite{qiu2022global}                                             & 15.05                                              & 13.27                                            & 69.61                                             & 21.95                  & 12.30                   & 12.93                   & 67.78                   & 13.81                   & 20.25                   & 19.13                   & 66.33                   & 42.77                   \\
					\midrule
			\multicolumn{13}{l}{\textit{Low-rank and Sparse Decomposition Methods}}                                                                                                                                                                                                                                                                                                                                                                                   \\
					\midrule
			IPI   \cite{gao2013infrared}                                                  & 25.62                                              & 31.28                                            & 80.24                                             & 11.97                  & 26.81                   & 28.93                   & 78.05                   & 26.17                   & 24.17                   & 33.27                   & 73.24                   & 43.39                   \\
			RIPT   \cite{dai2017reweighted}                                               & 11.38                                              & 14.32                                            & 76.33                                             & 22.66                  & 14.13                   & 16.74                   & 76.40                   & 28.34                   & 23.18                   & 32.63                   & 75.43                   & 56.21                   \\
			PSTNN    \cite{zhang2019infrared}                                               & 21.82                                              & 22.57                                            & 74.29                                             & 30.00                  & 27.69                   & 27.94                   & 79.41                   & 39.46                   & 20.41                   & 23.26                   & 69.89                   & 61.24                   \\
					\midrule
			\multicolumn{13}{l}{\textit{Deep Learning Methods}}                                                                                                                                                                                                                                                                                                                                                                                                       \\
					\midrule
			ACM   \cite{dai2021asymmetric}                                                & 70.47                                              & 71.71                                            & 92.26                                             & 9.94                   & 57.99                   & 57.46                   & 88.20                   & 21.48                   & 70.93                   & 70.08                   & 94.03                   & 76.19                   \\
			AGPCNet   \cite{zhang2023attention}                                             & 74.77                                              & 72.26                                            & \textcolor{red}{95.00}                            & 34.94                  & 65.59                   & 63.56                   & 90.55                   & 15.54                   & 85.04                   & 84.59                   & 93.54                   & 34.07                   \\
			DNANet     \cite{li2022dense}                                            & 75.70                                              & \textcolor{blue}{76.57}                          & 94.73                                             & 13.37                  & 67.19                   & 66.60                   & 88.38                   & 23.18                   & 87.70                   & 86.53                   & 93.69                   & 30.69                   \\
			ISNet      \cite{zhang2022isnet}                                           & 77.63                                              & 75.44                                            & 94.86                                             & 4.93                   & 66.02                   & 59.53                   & \textcolor{blue}{92.00} & 14.96                   & 84.53                   & 85.26                   & 89.45                   & 28.22                   \\
			UIUNet    \cite{wu2022uiu}                                             & 76.60                                              & 75.62                                            & 94.79                                             & 4.41                   & 62.61                   & 62.01                   & 88.67                   & 13.45                   & 85.34                   & 84.62                   & 94.50                   & 23.16                   \\
			MTUNet    \cite{wu2023mtu}                                             & 75.30                                              & 75.96                                            & 94.14                                             & 8.65                   & 64.39                   & 65.54                   & 89.49                   & 12.86                   & 83.93                   & 83.89                   & 94.27                   & 29.42                   \\
			SCTransNet   \cite{yuan2024sctransnet}                                        & 77.67                                              & 76.30                                            & 92.66                                             & 6.48                   & 68.39                   & \textcolor{blue}{67.34} & 90.71                   & 13.94                   & 88.73                   & 88.66                   & 93.84                   & 28.75                   \\
			SeRankDet   \cite{dai2024pick}                                          & \textcolor{blue}{78.04}                            & 76.51                                            & 94.82                                             & \textcolor{red}{2.77}  & \textcolor{blue}{71.50} & 67.26                   & 91.36                   & \textcolor{blue}{10.64} & \textcolor{blue}{90.43} & \textcolor{blue}{89.64} & \textcolor{blue}{94.81} & \textcolor{blue}{22.35} \\
					\midrule
			\rowcolor[gray]{0.9} $\star$ \textbf{DCGANet (Ours)} & \textcolor{red}{78.78}                             & \textcolor{red}{76.65}                           & \textcolor{blue}{94.96}                           & \textcolor{blue}{2.99} & \textcolor{red}{72.60}  & \textcolor{red}{69.31}  & \textcolor{red}{92.82}  & \textcolor{red}{9.20}   & \textcolor{red}{91.22}  & \textcolor{red}{90.72}  & \textcolor{red}{94.84}  & \textcolor{red}{20.34}  \\
			\bottomrule[1.5pt]
		\end{tabular}}
\end{table*}

\subsection{Visualization Analysis}
\begin{figure*}[htbp]
	\centering
	\includegraphics[width=1.0\linewidth]{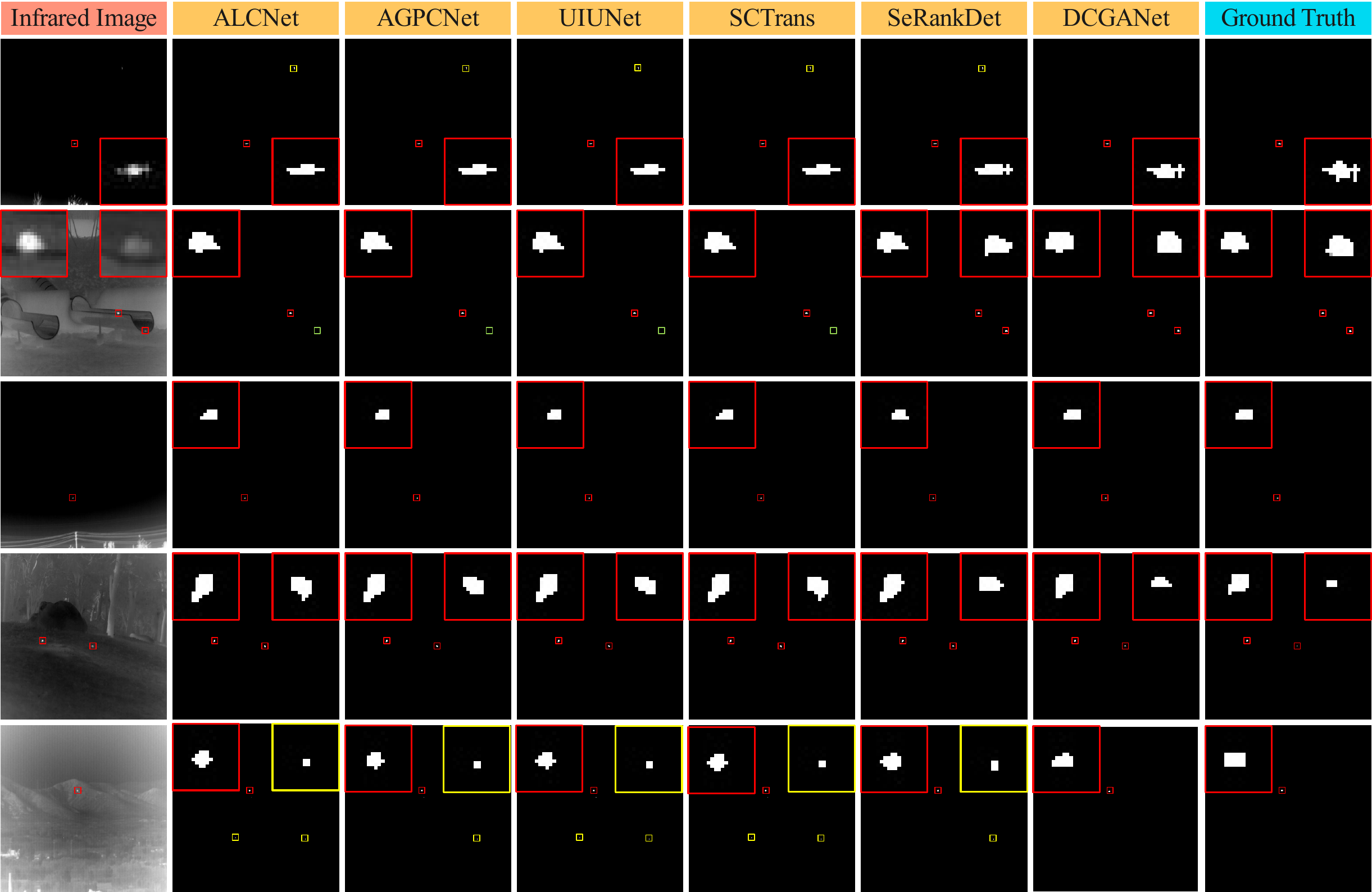}
	\caption{Visualization comparison of detection results on representative images from the three datasets, showcasing interference from land, oceans, and cityscapes. The \textcolor{red}{red}, \textcolor{green}{green}, and \textcolor{yellow}{yellow} boxes denote correct detections, false alarms, and missed detections, respectively.}
	\label{fig:visual1}
\end{figure*}

\begin{figure*}[htbp]
	\centering
	\includegraphics[width=1.0\linewidth]{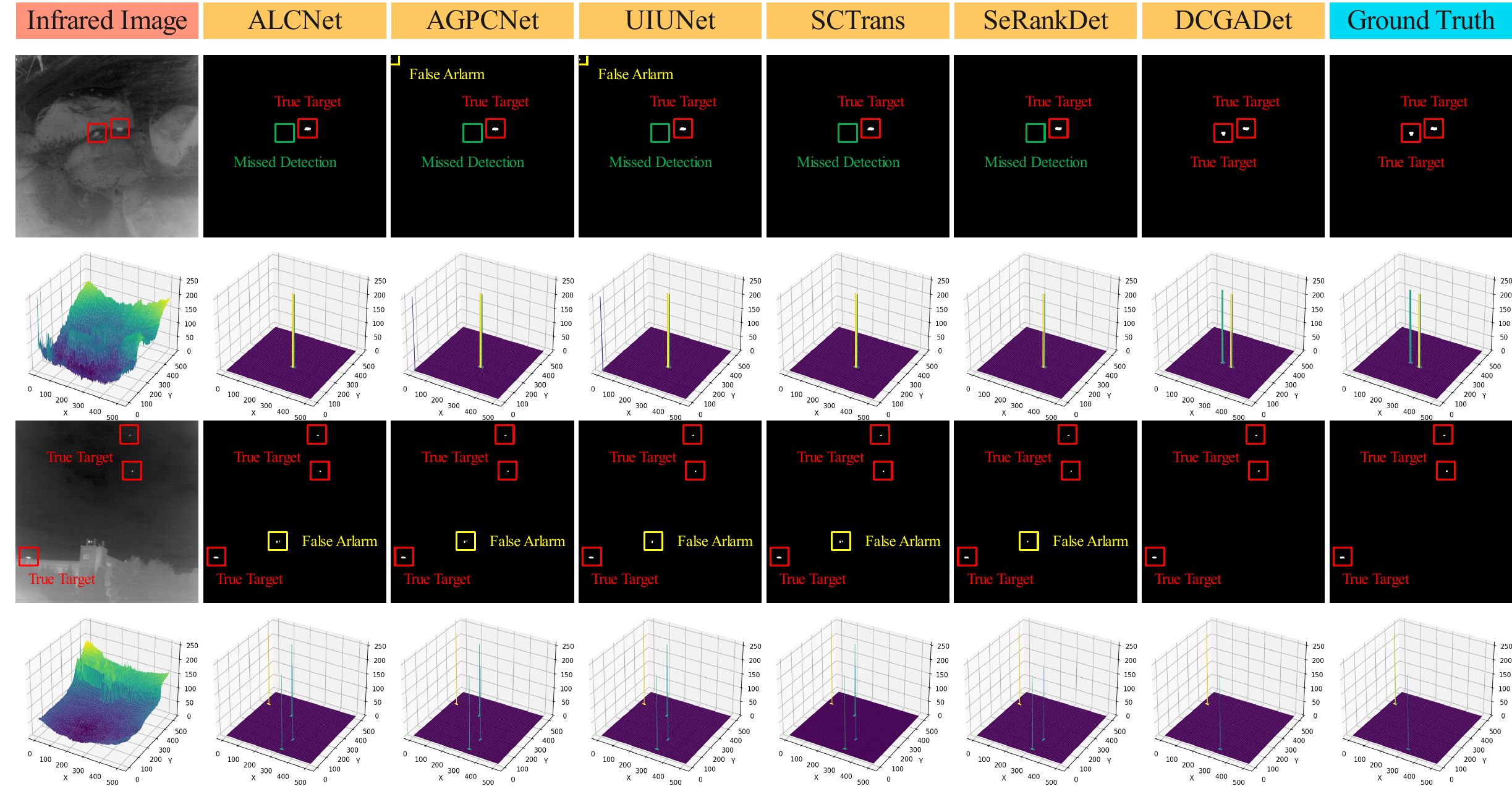}
	\caption{Further visualization comparisons. The main images show detection results, while the 3D mesh plots below each one highlight the intensity profile of the small target feature against its local background.}
	\label{fig:visual2}
\end{figure*}

To qualitatively validate the advantages of DCGANet, we analyze representative samples from the three datasets (Figures~\ref{fig:visual1} and~\ref{fig:visual2}). Traditional model-driven methods, which rely on handcrafted features, exhibit high false alarm rates (yellow boxes) and frequent missed detections. In contrast, deep learning-based methods leverage learned hierarchical representations to distinguish targets from complex background clutter more effectively, showcasing superior detection robustness.

\begin{figure*}
	\centering
	\includegraphics[width=1.0\linewidth]{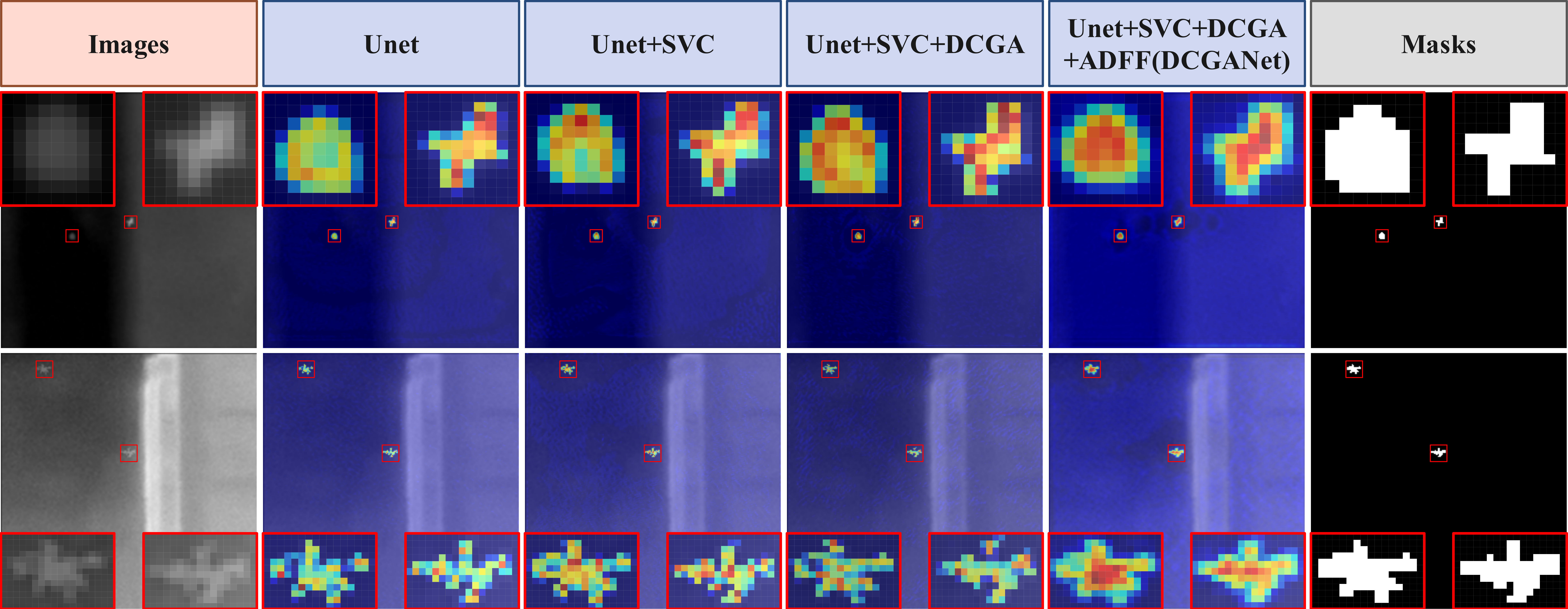}
	\caption{Qualitative comparison of attention maps on infrared images. From left to right: original images, predictions by U-Net (baseline), U-Net+SVC, U-Net+SVC+DCGA, U-Net+SVC+DCGA+ADFF (our full DCGANet), and ground-truth masks. The improvements in target localization and detail suppression of noise are evident with the addition of each module.}
	\label{fig:gradcam}
\end{figure*}

Gradient-weighted Class Activation Mapping (GradCAM) visualizations for DCGANet (Figure~\ref{fig:gradcam}) highlight the focus area for each module during target extraction.
The SVC module employs diverse convolutional forms, enabling robust feature preservation and edge refinement compared to the baseline.
The DCGA module adaptively refines the distinction between target and background, demonstrating enhanced precision in focusing on true targets.
Finally, the ADFF module improves edge preservation by dynamically fusing features, achieving sub-pixel accuracy in localization. These results collectively demonstrate that DCGANet's modules operate synergistically, with each component exhibiting clear and discriminative attention patterns.

\subsection{Ablation Study}

\subsubsection{Component-wise Performance Analysis}
Ablation experiments were conducted to systematically evaluate the proposed SVC, DCGA, and ADFF modules, demonstrating their individual and integrated impacts on model performance. As shown in Table~\ref{tab:Model-wise}, the baseline U-Net was used for comparison.

\begin{enumerate}
	\item \textbf{Independent Efficacy}: Each module individually yields distinct performance improvements over the baseline (a), corroborating its theoretical motivation. For instance, adding DCGA alone (c) improves IoU by 7.67\% on IRSTD-1K.

	\item \textbf{Effectiveness of SVC}: Replacing standard convolutions with our SVC module (b) elevates the IoU by 4.53\% on NUAA-SIRST and 7.35\% on IRSTD-1K, proving its superiority in encoding multi-scale target features while suppressing clutter.

	\item \textbf{Synergy of Modules}: Combining SVC and DCGA (d) provides a significant boost over using either module alone. The full model, DCGANet (e), which also includes ADFF, achieves the overall best performance, demonstrating the powerful synergy of the three proposed components.
\end{enumerate}

Furthermore, Figure~\ref{fig:gradcam} presents a qualitative comparison of attention maps under different model configurations. The visualizations showcase the incremental benefits of each module in refining target localization and enhancing feature representation.

\begin{table}[h]
	\centering
	\small
	\caption{Quantitative ablation study of DCGANet's core components. We measure the performance improvement by sequentially adding Selective Variable Convolution (SVC), Dynamic Context-Guided Attention (DCGA), and Adaptive Dynamic Feature Fusion (ADFF) to a U-Net baseline.}
	\label{tab:Model-wise}
    	\resizebox{\columnwidth}{!}{

	\begin{tabular}{c|ccc|cc|cc|cc}
		\toprule[1.5pt]
		\multirow{2}{*}{\textbf{Strategy}} & \multicolumn{3}{c|}{\textbf{Modules}} & \multicolumn{2}{c|}{\textbf{NUAA-SIRST}} & \multicolumn{2}{c|}{\textbf{IRSTD-1K}} & \multicolumn{2}{c}{\textbf{NUDT-SIRST}}                                                                                      \\
		                                   & \textbf{SVC}                          & \textbf{DCGA}                            & \textbf{ADFF}                          & IoU$\uparrow$                           & nIoU$\uparrow$ & IoU$\uparrow$  & nIoU$\uparrow$ & IoU$\uparrow$  & nIoU$\uparrow$ \\
		\midrule
		(a)                                & \ding{55}                             & \ding{55}                                & \ding{55}                              & 72.15                                   & 72.36          & 61.87          & 60.06          & 86.25          & 86.03          \\
		(b)                                & \checkmark                            & \ding{55}                                & \ding{55}                              & 76.59                                   & 75.42          & 69.07          & 66.76          & 88.76          & 88.73          \\
		(c)                                & \ding{55}                             & \checkmark                               & \ding{55}                              & 76.94                                   & 75.64          & 69.39          & 67.43          & 89.34          & 89.20          \\
		(d)                                & \checkmark                            & \checkmark                               & \ding{55}                              & 78.12                                   & 75.88          & 71.89          & 69.55          & 90.26          & 89.82          \\
		\rowcolor[gray]{0.9}
		(e)                                & \checkmark                            & \checkmark                               & \checkmark                             & \textbf{78.78}                          & \textbf{76.65} & \textbf{72.60} & \textbf{69.31} & \textbf{91.22} & \textbf{90.72} \\
		\bottomrule[1.5pt]
	\end{tabular}}
\end{table}

\subsubsection{Ablation of SVC Module Design}
The SVC module integrates Standard Convolution (SConv), Deformable Convolution (DConv), and Multiple Dilation Convolution (MDConv). To analyze each branch's contribution, we conduct an ablation study shown in Table~\ref{tab:SVC_ablation}. The results reveal that while adding DConv provides a moderate performance gain (c), the addition of MDConv yields significantly greater improvements (b). The full three-branch SVC (d) achieves the best performance, demonstrating that multi-rate dilated convolutions play a pivotal role in expanding the receptive field and capturing crucial hierarchical context.

\begin{table}[h]
	\centering
	\small
	\caption{Ablation study of the convolutional branches within the SVC module. We start with a standard convolution (SConv) and add the other branches. Results on NUDT-SIRST are included for comprehensive analysis.}
	\label{tab:SVC_ablation}
    	\resizebox{\columnwidth}{!}{

	\begin{tabular}{c|ccc|cc|cc|cc}
		\toprule[1.5pt]
		\multirow{2}{*}{\textbf{Strategy}} & \multicolumn{3}{c|}{\textbf{Branches}} & \multicolumn{2}{c|}{\textbf{NUAA-SIRST}} & \multicolumn{2}{c|}{\textbf{IRSTD-1K}} & \multicolumn{2}{c}{\textbf{NUDT-SIRST}}                                                                                      \\
		                                   & SConv                                  & DConv                                    & MDConv                                 & IoU$\uparrow$                           & nIoU$\uparrow$ & IoU$\uparrow$  & nIoU$\uparrow$ & IoU$\uparrow$  & nIoU$\uparrow$ \\
		\midrule
		(a)                                & \checkmark                             & \ding{55}                                & \ding{55}                              & 77.17                                   & 74.52          & 69.29          & 67.35          & 88.35          & 88.18          \\
		(b)                                & \checkmark                             & \ding{55}                                & \checkmark                             & 78.49                                   & 75.36          & 71.08          & 68.04          & 90.46          & 90.31          \\
		(c)                                & \checkmark                             & \checkmark                               & \ding{55}                              & 77.76                                   & 75.13          & 70.54          & 67.57          & 89.25          & 89.03          \\
		\rowcolor[gray]{0.9}
		(d)                                & \checkmark                             & \checkmark                               & \checkmark                             & \textbf{78.78}                          & \textbf{76.65} & \textbf{72.60} & \textbf{69.31} & \textbf{91.22} & \textbf{90.72} \\
		\bottomrule[1.5pt]
	\end{tabular}}
\end{table}

\subsubsection{Ablation of DCGA Module Design}
We compare DCGA against other mainstream attention mechanisms (SE, DAM, CBAM) in Table~\ref{tab:DCGA_comparison}. DCGA significantly outperforms them, showing its ability to better recalibrate feature distributions by learning channel-specific Spatial Importance Maps (SIMs), which is critical for enhancing infrared small target features.

\begin{table}[htbp]
	\centering
	\small
	\caption{Comparison of DCGA with other mainstream attention mechanisms.}
	\label{tab:DCGA_comparison}
    	\resizebox{\columnwidth}{!}{
	\begin{tabular}{l|cc|cc|cc}
		\toprule[1.5pt]
		\multirow{2}{*}{\textbf{Strategy}} & \multicolumn{2}{c|}{\textbf{NUAA-SIRST}} & \multicolumn{2}{c|}{\textbf{IRSTD-1K}} & \multicolumn{2}{c}{\textbf{NUDT-SIRST}}                                                    \\
		                                   & IoU$\uparrow$                            & nIoU$\uparrow$                         & IoU$\uparrow$                           & nIoU$\uparrow$ & IoU$\uparrow$  & nIoU$\uparrow$ \\
		\midrule
		SE                                 & 75.63                                    & 74.81                                  & 65.43                                   & 65.76          & 87.33          & 87.38          \\
		DAM                                & 76.71                                    & 74.92                                  & 67.41                                   & 66.38          & 87.67          & 88.02          \\
		CBAM                               & 76.83                                    & 74.79                                  & 68.38                                   & 65.88          & 88.64          & 88.40          \\
		\rowcolor[gray]{0.9}
		\textbf{DCGA (Ours)}               & \textbf{78.78}                           & \textbf{76.65}                         & \textbf{72.60}                          & \textbf{69.31} & \textbf{91.22} & \textbf{90.72} \\
		\bottomrule[1.5pt]
	\end{tabular}}
\end{table}

Furthermore, we conducted a detailed ablation of the DCGA's internal design choices (Table~\ref{tab:DCGA_internal_ablation}). The study validates our design:
\begin{itemize}
	\item \textbf{DCGA\textsubscript{cas}}: A cascaded connection of Channel (CA) and Spatial Attention (SA) performs slightly worse than our parallel design, suggesting parallel processing is more effective for this task.
	\item \textbf{DCGA\textsubscript{w/o r}}: Removing the refinement stage entirely causes a significant performance drop, confirming the importance of the refinement process.
	\item \textbf{DCGA\textsubscript{w/o CS}}: Removing only the Channel Shuffle (CS) from the refinement stage also degrades performance, indicating that CS is crucial for cross-channel information interaction.
\end{itemize}

\begin{table}[htbp]
	\centering
	\small
	\caption{Detailed ablation study on the internal design of the DCGA module. We compare our full design against three variants: a cascaded version (DCGA\textsubscript{cas}), one without the refinement stage (DCGA\textsubscript{w/o r}), and one without the channel shuffle operation (DCGA\textsubscript{w/o CS}). The best results are highlighted in \textbf{bold}.}
	\label{tab:DCGA_internal_ablation}
    	\resizebox{\columnwidth}{!}{
	\begin{tabular}{l|l|cccc}
		\toprule[1.5pt]
		\textbf{Dataset}            & \textbf{Metric}      & \textbf{DCGA\textsubscript{cas}} & \textbf{DCGA\textsubscript{w/o r}} & \textbf{DCGA\textsubscript{w/o CS}} & \textbf{DCGA (Ours)} \\
		\midrule
		\multirow{2}{*}{NUAA-SIRST} & IoU (\%) $\uparrow$  & 76.95                            & 72.18                              & 75.43                               & \textbf{78.78}       \\
		                            & nIoU (\%) $\uparrow$ & 75.60                            & 72.26                              & 74.07                               & \textbf{76.65}       \\
		\midrule
		\multirow{2}{*}{IRSTD-1K}   & IoU (\%) $\uparrow$  & 69.39                            & 61.86                              & 68.51                               & \textbf{72.60}       \\
		                            & nIoU (\%) $\uparrow$ & 67.42                            & 60.48                              & 66.31                               & \textbf{69.31}       \\
		\midrule
		\multirow{2}{*}{NUDT-SIRST} & IoU (\%) $\uparrow$  & 89.36                            & 85.61                              & 87.58                               & \textbf{91.22}       \\
		                            & nIoU (\%) $\uparrow$ & 89.10                            & 85.89                              & 87.41                               & \textbf{90.72}       \\
		\bottomrule[1.5pt]
	\end{tabular}}
\end{table}

\section{Conclusion} \label{sec:conclusion}

In this paper, we propose a dynamic content-guided attention multiscale feature aggregation network (DCGANet), which adopts a "coarse-to-fine" attention principle to achieve high detection accuracy. First, the selective variable convolution (SVC) module is introduced, integrating standard convolution, deformable convolution, and multi-rate dilated convolution. This module expands the receptive field and enhances non-local features, thereby effectively improving the discrimination between targets and backgrounds. Second, the core component of DCGANet is the Dynamic Content-Guided Attention (DCGA) module, which uses two-stage attention to first direct the network's focus to significant regions in feature maps and then determine whether these regions correspond to targets or background interference. By retaining the most salient responses, this mechanism suppresses false alarms effectively. Additionally, the Adaptive Dynamic Feature Fusion (ADFF) module is proposed to substitute for static feature cascading. The dynamic feature fusion strategy enables DCGANet to adaptively integrate contextual features, enhancing its capability to discriminate between true targets and false alarms. DCGANet establishes new benchmarks across multiple datasets.

\bibliographystyle{IEEEtran}
\bibliography{./reference.bib}

% 作者介绍

\end{document}